\documentclass[12pt]{iopart}

\usepackage[pdftex]{graphicx}
\usepackage{iopams}
\usepackage{epstopdf}
\usepackage{graphicx}
\usepackage{cite}
\usepackage{gensymb}

\begin{document}
\def\be{\begin{equation}}
\def\ee{\end{equation}}
\def\bea{\begin{eqnarray}}
\def\eea{\end{eqnarray}}
\def\rp{r_{+}}
\def\rmm{r_{-}}

\title{
\bf The Dual Gonihedric 3D Ising Model
}
\date{March 2011}
\author{D. A. Johnston}
\address{Dept. of Mathematics, Heriot-Watt University,
Riccarton,Edinburgh, EH14 4AS, Scotland}

\author{R. P. K. C. M. Ranasinghe}
\address{Department of Mathematics, University of Sri Jayewardenepura,
Gangodawila, Sri Lanka.}


\begin{abstract}

We investigate the dual of the $\kappa=0$ Gonihedric Ising model
on a $3D$ cubic lattice, 
which may be written as an anisotropically coupled Ashkin-Teller model. 
The original $\kappa=0$ Gonihedric model has a purely plaquette interaction, displays a first order transition and possesses a highly degenerate ground state. 

We find that the dual model  admits a similar large ground state degeneracy
as a result of the anisotropic couplings and investigate the 
coupled mean field equations for the model on a single cube.
We also carry out Monte Carlo simulations which confirm a first order phase transition in the model and suggest that
the ground state degeneracy persists throughout the low temperature phase. Some exploratory cooling simulations also hint at non-trivial dynamical behaviour.

\end{abstract} 

\maketitle


\section{Introduction}

\begin{figure}[h]
\centering
\includegraphics[height=3.5cm]{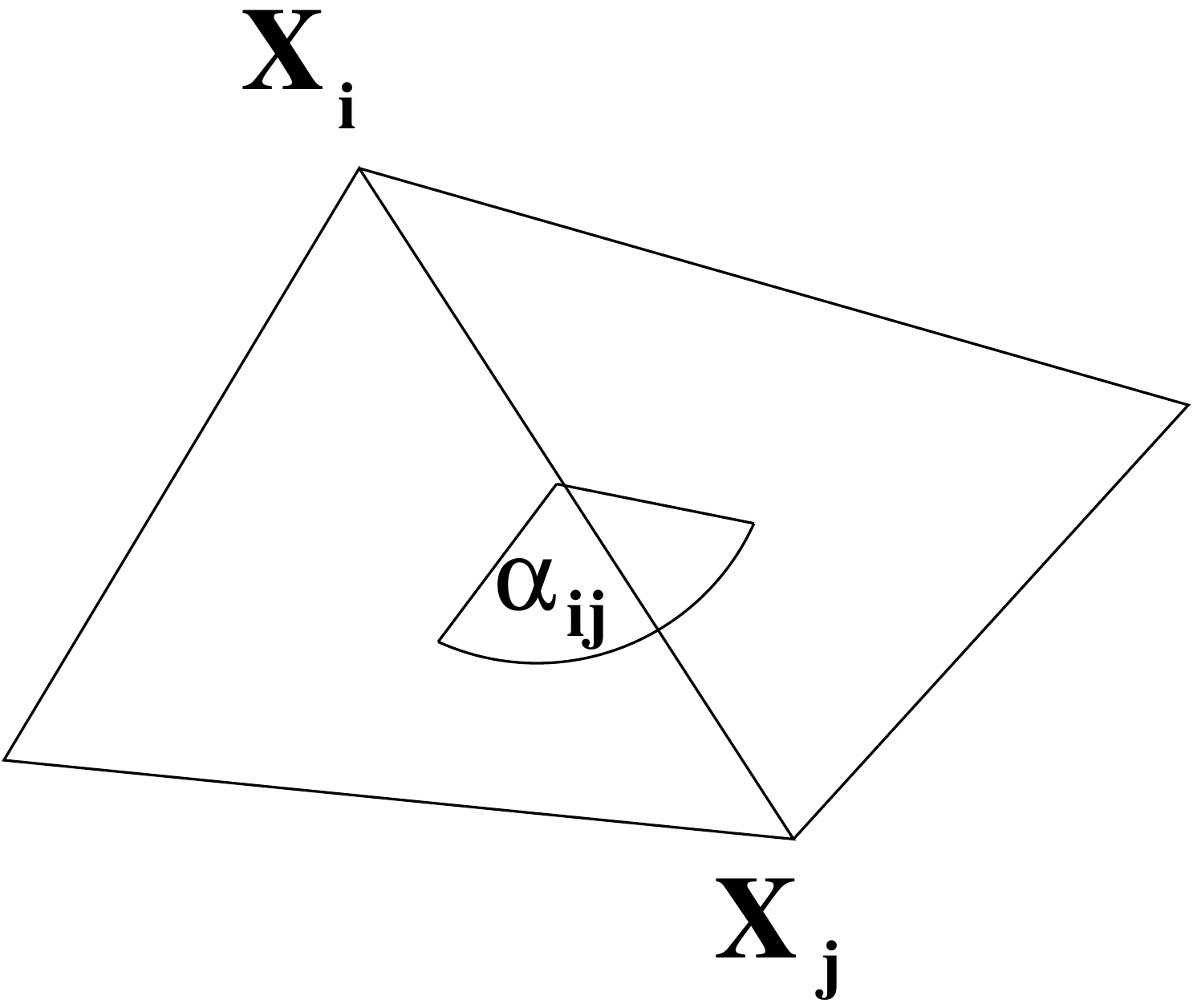}
\caption{Two adjacent triangles in a triangulation of a 
surface with a common edge $\langle ij \rangle$, showing 
the co-ordinates of the endpoints $\vec X_{i,j}$ and the dihedral angle $\alpha_{ij}$}
\label{dihedral}       
\end{figure}

The Gonihedric Ising model has an interesting history, having originally been formulated as 
a fixed lattice version of a discretized string/triangulated random surface action suggested by Savvidy {\it et.al.},
whose action was given by
\cite{1}
\begin{equation}
S = {1 \over 2} \sum_{\langle ij \rangle} | \vec X_i - \vec X_j | \; \theta (\alpha_{ij}),
\label{steiner}
\end{equation}
where
$\theta(\alpha_{ij}) = | \pi - \alpha_{ij} |$
and $\alpha_{ij}$ is the dihedral angle between the
embedded neighbouring triangles with a common link $\langle ij \rangle$.
The  $| \vec X_i - \vec X_j |$ are the lengths of the embedded triangle edges as shown in Fig.~(\ref{dihedral}). 
The aim of this action was to weight the edges of non-coplanar adjoining triangles on a discretized surface, rather than the triangle areas as is the case with a Gaussian action, in an attempt to search for a continuum limit which might be related to a string theory. The discretized random surfaces formed from gluing together triangles were intended to model Euclidean string worldsheets.

Translating this action
onto a fixed cubic lattice and asking that the requisite surfaces be represented by the plaquettes of spin cluster boundaries in some Ising-like model trivializes the edge length $| \vec  X_i - \vec X_j |$ dependence, so the statistical weights of surface configurations depend solely on
the $\theta(\alpha_{ij}) = | \pi - \alpha_{ij} |$ factors, where the $\alpha_{ij}$ are now restricted to multiples of $\pi / 2$ radians. The statistical weights of such plaquette surface configurations will be determined entirely by the number of bends and self-intersections they contain.

The mapping between the energies of a gas of plaquette surfaces and generalised Ising models was studied in some detail by Cappi {\it et.al.} \cite{2} who calculated the energies of spin cluster boundaries on the 3D cubic lattice for an Ising spin ($\pm 1$) Hamiltonian which contains nearest neighbour $\langle i,j\rangle $,
next to nearest neighbour $\langle \langle i,j\rangle \rangle $ and plaquette $[i,j,k,l]$
terms
\begin{equation}
-\beta H = J_1  \sum_{\langle ij\rangle }\sigma_{i} \sigma_{j} +
 J_2 \sum_{\langle \langle i,j\rangle \rangle }\sigma_{i} \sigma_{j} 
+ J_3 \sum_{[i,j,k,l]}\sigma_{i} \sigma_{j}\sigma_{k} \sigma_{l}.
\label{e0}
\end{equation}
The couplings in such models can be related to the couplings for the area energy of plaquettes in  
spin cluster boundaries, $\beta_A$, the energy cost of a right-angled
bend between two such adjacent plaquettes, $\beta_C$, and the energy cost, $\beta_I$, for the intersection of four plaquettes having a link in common 
\begin{eqnarray}
\beta_A &=&  2 J_1 + 8 J_2 \nonumber \\
\beta_C &=&  2 J_3 - 2 J_2  \nonumber \\
\beta_I &=&  -4 J_2 - 4 J_3  \; .
\end{eqnarray}
The original 3D Gonihedric model \cite{3} constitutes
a particular one-parameter slice of this family of Hamiltonians:
\begin{equation}
H = - 2 \kappa \sum_{\langle ij\rangle }\sigma_{i} \sigma_{j}  +
\frac{\kappa}{2}\sum_{\langle \langle i,j\rangle \rangle }\sigma_{i} \sigma_{j} 
- \frac{1-\kappa}{2}\sum_{[i,j,k,l]}\sigma_{i} \sigma_{j}\sigma_{k} \sigma_{l}.
\label{e1}
\end{equation}
For this ratio of couplings $\beta_A=0$, which means that the edges and intersections
of spin cluster boundaries are weighted rather than their area, which is the antithesis of the usual
3D Ising model with only nearest neighbour spin interactions where $\beta_I = \beta_C = 0$.
The energy of the spin cluster boundaries for the Gonihedric model on a cubic
lattice is simply given by
$
E=n_2 + 4 \kappa n_4 \, ,
$
where $n_2$ is the number
of links where two plaquettes on a spin cluster boundary meet at a right angle,
$n_4$ is the number of links where four plaquettes
meet at right angles and $\kappa$ is the free parameter. It is worth remarking that the language we have employed
implicitly assumes that spin cluster boundaries can be clearly identified. As we shall see below this may not be such a simple matter for the dual Gonihedric model (it is similarly complicated for the original Gonihedric action, at least when $\kappa=0$).

When   $\kappa=0$ the Gonihedric Hamiltonian becomes
a purely plaquette term 
\begin{equation}
H =  - \frac{1}{2} \sum_{[i,j,k,l]}\sigma_{i} \sigma_{j}\sigma_{k} \sigma_{l}
\label{e2}
\end{equation}
which is {\it not} the 3D gauge Ising model, since the spins live on the vertices
rather than the edges of the lattice. This plaquette action displays a first order transition
surrounded by a region of metastability \cite{4}. It also displays interesting dynamical behaviour with a dynamical transition at the lower boundary of the metastable region which appears to have many glassy characteristics \cite{5a,5,5e,5b}. This is intriguing because there is no quenched disorder in the Hamiltonian.

The dual to the $\kappa=0$ Gonihedric Ising model
was constructed by Savvidy and Wegner \cite{6}. They considered a high temperature expansion of the partition function for the Hamiltonian in equ.(\ref{e2})
\begin{eqnarray}
Z (\beta)  &=& \sum_{\{\sigma\}}  \exp (- \beta H ) \nonumber \\
&=& \sum_{\{\sigma\}} \prod_{[i,j,k,l]} \cosh  \left(\frac{\beta}{2} \right) \left[1 + \tanh \left(\frac{\beta}{2} \right) ( \sigma_i \sigma_j \sigma_k \sigma_l )\right] 
\label{z2}
\end{eqnarray}
which can be written as
\begin{equation}
Z (\beta)  = \left[ 2 \cosh \left(\frac{\beta}{2} \right) \right]^{3 L^3} \sum_{\{ S \}} \left[\tanh \left(\frac{\beta}{2} \right)\right]^{n ( S ) } 
\label{z2a}
\end{equation}
on an $L^3$ cubic lattice.
The sum runs over closed surfaces with an even number of plaquettes at any vertex and $ n ( S )$ is the number of 
plaquettes in a given surface. This ensemble can be constructed from three differently oriented elementary ``matchbox'' surfaces of the form
shown in Fig.~(2) along with the unshaded cube.
A spin variable representing each matchbox then sits at the centre of the cube on the dual lattice 
and any surface in the ensemble can be constructed as a product of the elementary matchboxes and unshaded cubes. When two shaded 
matchbox faces  overlap they annihilate to give an unshaded face, so the shaded faces can be thought of as carrying a negative sign.

The low
temperature expansion  of the dual Hamiltonian
\begin{equation}
H_{dual} = - \frac{1}{2}\sum_{\langle ij \rangle} \sigma_{i}  \sigma_{j} 
- \frac{1}{2} \sum_{\langle ik \rangle } \tau_{i}  \tau_{k} 
- \frac{1}{2} \sum_{ \langle jk \rangle} \eta_{j} \eta_{k} 
\label{dual1}
\end{equation}
gives precisely this structure, where $\sigma, \tau$ and $\eta$ represent each of the 
possible matchbox orientations.
In equ.(\ref{dual1})
the spins $\sigma, \tau$
and $\eta$ live on the vertices
of the dual lattice 
and the sums are along its orthogonal edges $ij, ik$ and $jk$.
\begin{figure}[h]
\begin{center}
\includegraphics[height=4cm]{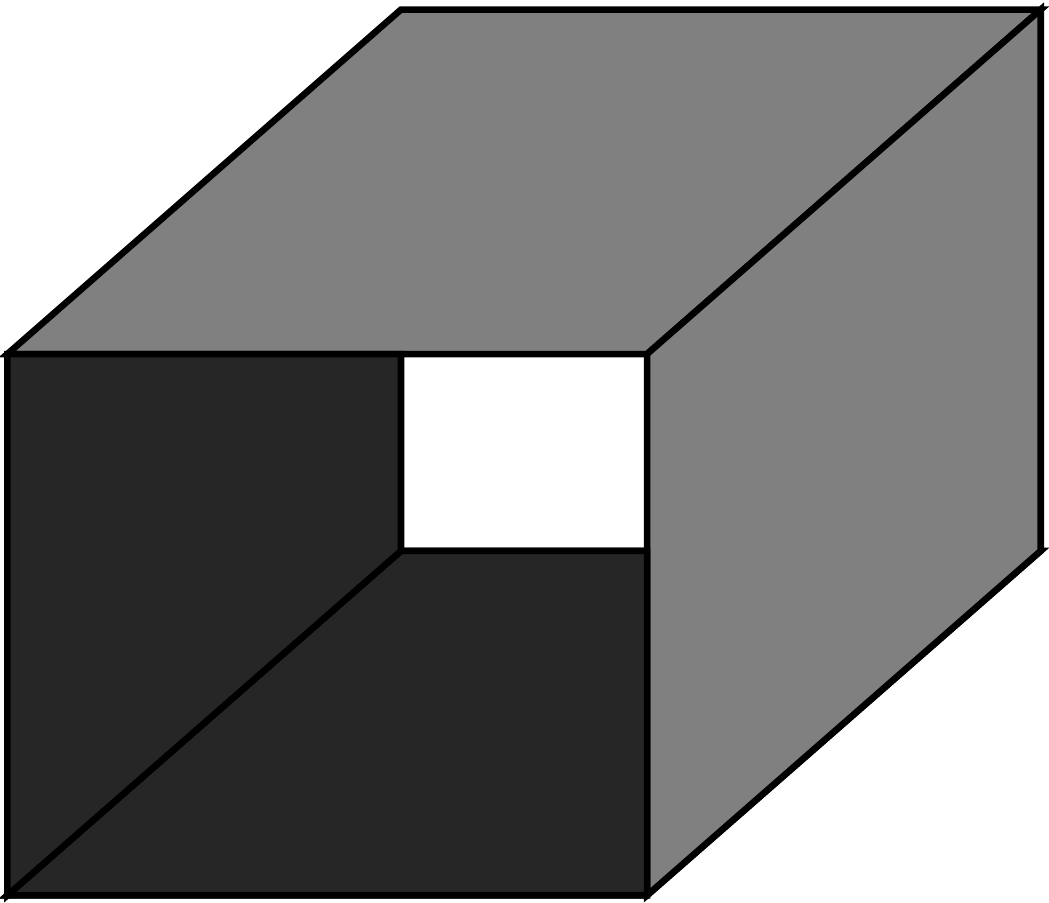}
\label{matchbox} 
\caption{One of the three possible orientations of an elementary matchbox surface.}
\end{center}
\end{figure}
They satisfy
\begin{eqnarray}
e \sigma &=&  \sigma \; , \;  \; e \tau = \tau \; ,  \; \; e \eta = \eta \nonumber \\	
\sigma^2 &=& \tau^2 = \eta^2 = e \\
\sigma \tau &=& \eta \, , \; \; \tau \eta = \sigma \; , \; \; \eta \sigma = \tau \nonumber
\end{eqnarray}
with $e$ representing the unshaded cube,
which means the spins live in the fourth order Abelian group.
For convenience in simulations the spins may also be considered as Ising ($\pm 1$) spins
if we set $\eta_i = \sigma_i \; \tau_i $,
which recasts the Hamiltonian into 
\begin{equation} 
H_{dual} = - \frac{1}{2} \sum_{ \langle ij \rangle} \sigma_{i}  \sigma_{j} 
- \frac{1}{2}  \sum_{ \langle ik \rangle } \tau_{i}  \tau_{k} 
-  \frac{1}{2} \sum_{\langle jk \rangle} \sigma_{j} \sigma_{k} \tau_{j}  \tau_{k} \, ,
\label{dual2}
\end{equation}
which is recognizable as an anisotropically coupled Ashkin-Teller \cite{7} model with equal couplings. Without 
the four-spin term this would simply be two uncoupled 1D Ising chains arranged in perpendicular
directions and thus display no transition(s), but as we see below the coupling via the 
Ashkin-Teller energy term in the third direction gives non-trivial behaviour.

In the isotropically 
coupled case the ratio in equ.(\ref{dual2}) corresponds to the increased symmetry point
of the standard Ashkin-Teller model
where the generic $\mathbb{Z}_2 \times \mathbb{Z}_2$ symmetry is promoted to $\mathbb{Z}_4$.
This can be seen explicitly by rewriting the Hamiltonian in terms of the four double spins 
$S_i = ( \pm 1, \pm 1)$ to give the 4-state Potts Hamiltonian 
\begin{equation} 
H_{dual, isotropic} = - \frac{1}{2} \sum_{ \langle ij \rangle}  \left( 4 \delta_{S_i , S_j }  - 1 \right)
\label{4Potts}
\end{equation}
where the sum now runs over all the edges orientations.
4-state Potts critical behaviour (i.e. a first order phase transition in 3D \cite{8}) is thus found in the isotropic case.  

In the remainder of the paper we 
consider the behaviour of the 3D dual Gonihedric Hamiltonian,
as formulated in equ.(\ref{dual2}) as an anisotropic Ashkin-Teller model, in its own right. We first look at the ground state structure of the model, highlighting the similarities with the plaquette action 
and in the light of this discuss coupled mean field equations on a cube.
We then report on Monte Carlo simulations which are sufficient to confirm the nature of the phase transition and  look at 
the effect of different cooling rates on the low temperature behaviour.  
The ground state, mean field and Monte-Carlo investigations all highlight the difficult of formulating a standard magnetic order parameter, and we discuss the  implications.

\section{Ground State}

In the isotropically coupled Ashkin-Teller model with equal positive couplings four equivalent magnetized ground states are possible, with the ($\sigma$,$\tau$) spins taking the values ($ \pm, \pm $) at every site, so only paramagnetic or  ferromagnetic behaviour
is seen for the individual $\sigma$ and $\tau$ spins in this coupling regime.
To investigate the ground state/zero-temperature
structure of the dual Gonihedric model 
we use an approach which proved useful for  the original undualized Gonihedric model 
and consider possible spin configurations on an elementary cube \cite{2}. 
This is sufficiently large to capture non-trivial structure
and the full ground state is then obtained by tiling the 3D cubic lattice with compatible single cube configurations.

The full lattice
Hamiltonian may be written as a sum over the individual cube Hamiltonians $h_C$,
\begin{equation}
h_c =  -\frac{1}{8}\sum_{ \langle i,j \rangle} \sigma_{i} \sigma_{j} - \frac{1}{8}\sum_{ \langle i,k \rangle} \tau_{i} \tau_{k}
-   \frac{1}{8} \sum_{ \langle j,k \rangle} \sigma_{j} \sigma_{k} \tau_{j} \tau_{k} \, ,
\end{equation}
where the additional symmetry factor of $\frac{1}{4}$ takes account of one edge being shared by four cubes. 
If a configuration of spins on a cube minimizes $h_c$
the full lattice ground state energy density will simply be 
given by $h_c$. 

\begin{figure}[h]
\centering
\includegraphics[height=8cm]{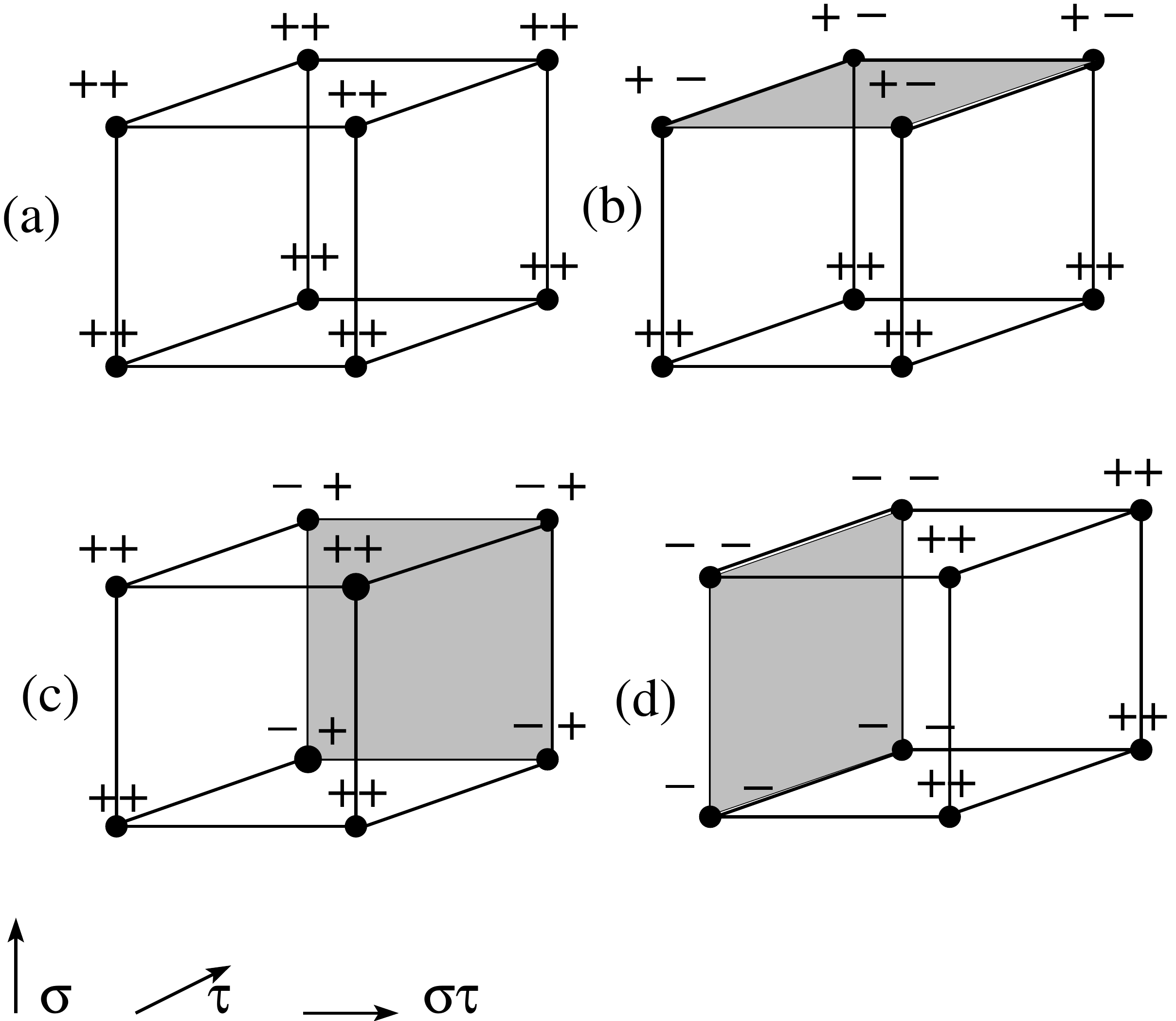}
\caption{Some possible ground state spin configurations on a cube, the $\sigma,\tau$ values are shown at each site. The directions of the couplings in the Hamiltonian are indicated, as are the faces on which spins are flipped. }
\label{ground} 
\end{figure}

Looking at the configurations in Fig.~(\ref{ground})
immediately makes it clear that there is considerably more freedom for possible ground states in the anisotropically coupled Hamiltonian of equ.(\ref{dual2}) than in the standard isotropic Ashkin-Teller model. 
In addition to the reference ferromagnetic ground  states in Fig.~(\ref{ground}a)
it is possible
to flip a single horizontal face  of $\tau$ spins on the cube at zero energy cost as in Fig.~(\ref{ground}b), or a vertical face of
either $\sigma$ (Fig.~(\ref{ground}c)) or both $\sigma$ and $\tau$ spins (Fig.~(\ref{ground}d)) on the differently oriented vertical faces. Flipping {\it two} faces in the same orientation takes one between the different possible ground states of the standard isotropic model, so it is the ability to flip a single face at zero energy cost which confers the extra freedom in the anisotropic model.

It is also possible
to combine the differently oriented single face flips on the cube  without increasing the energy of the configuration.
Tiling the entire lattice with such configurations then shows that ground states may contain flipped planes of $\sigma$, $\tau$
or $\sigma \tau$ spins (depending on the orientation) with respect to reference purely ferromagnetic ground states. It is possible for two orientations of flipped spin planes to intersect pairwise along a line or for three differently oriented planes to intersect at a point.
The distribution of flipped spin planes in a ground state is thus completely arbitrary.

The ground state degeneracy of the dual Gonihedric model is thus similar to that of the original plaquette Hamiltonian of equ.~(\ref{e2}) where (possibly orthogonal, intersecting) planes of  spins may also be flipped at zero energy cost. 
Intriguingly, this leads to the same difficulties in defining a suitable magnetic order parameter
for the dual Hamiltonian as one faces with the plaquette Hamiltonian. Since arbitrary, and arbitrarily separated, spin planes may be flipped at zero energy cost a standard, or even a staggered
magnetization, will generically be zero in such a state.

Of course, it is not guaranteed that ground state/zero-temperature degeneracies are maintained 
at finite temperature. Indeed, for the original Gonihedric model low temperature expansions
by Pietig and Wegner \cite{10} showed that when $\kappa \ne 0$, where it was still possible to flip 
arbitrary {\it parallel} spin planes to give a sandwich ground state, the ferromagnetic  state had a lower free energy at finite temperature. The degeneracy did, however, persist at finite temperatures for the $\kappa=0$ model. In the case of the dual Gonihedric model the Monte-Carlo simulations discussed below find $ \langle \sigma \rangle  = \langle \tau \rangle =  \langle \sigma \tau \rangle  \sim 0$ in the low temperature phase, though there is a clear sign of a phase transition in the energetic observables and also the magnetic susceptibilities. This suggests that the 
flip symmetry persists at finite temperature throughout the low temperature phase.

\section{Mean Field}

\begin{figure}[h]
\centering
\includegraphics[height=5cm]{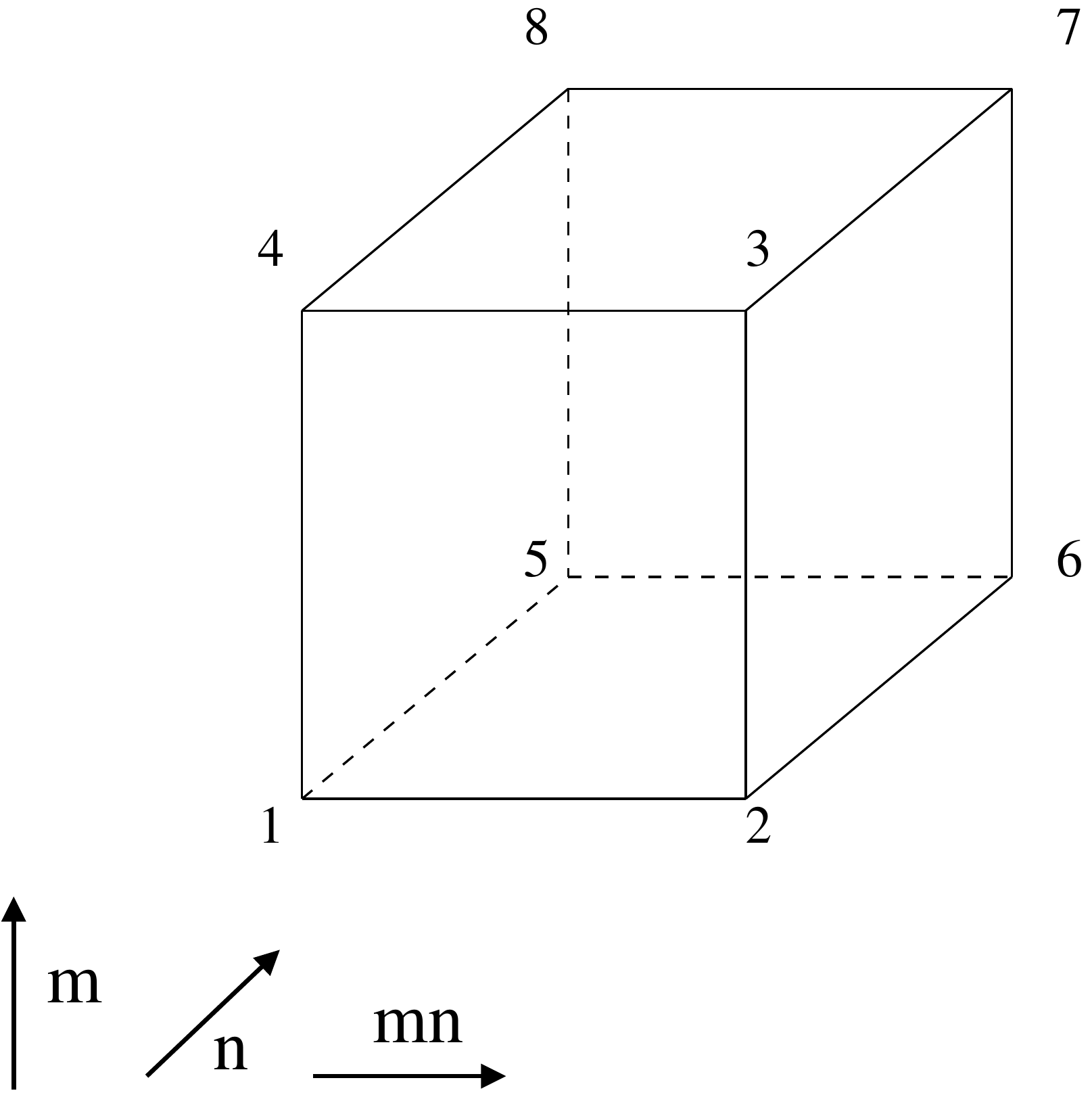}
\caption{The labelling of sites used in writing the mean field equations for the cube. The directions of the mean field spin couplings in the Hamiltonian are again indicated. }
\label{Fig1} 
\end{figure}

In the mean field approximation an expression for the free energy is written by replacing the exact values of the spins with average site magnetizations and adding an entropy term. To take account of non-trivial structure in such a calculation one can again work at the level of the cubes, as for the ground state,
and write down the coupled equations for the sixteen site magnetizations on the cube.   
The calculation 
of the mean field free energy in this manner is thus a direct elaboration of the
method used
to investigate the ground states.
The total mean field free energy is written as a sum of the elementary cube free energies $\phi(m_{C}, n_C)$, given by
\begin{eqnarray}
\beta \, \phi{(m_{C},n_C)} &=&  - \frac{\beta}{8}\sum_{ \langle i,j \rangle \subset C} m_{i} m_{j} -  \frac{\beta}{8}\sum_{ \langle i,k \rangle \subset C} n_{i} n_{k}
- \frac{\beta}{8}\sum_{\langle j,k \rangle \subset C} m_{j} m_{k} n_j n_k \nonumber \\
&+& \frac{1}{16}
\sum_{i \subset C}[(1+m_{i}) \ln(1+m_{i}) + (1- m_{i}) \ln(1 - m_{i})] \nonumber \\
&+& \frac{1}{16}
\sum_{i \subset C}[(1+n_{i}) \, \ln(1+n_{i}) \, + \, (1 - n_{i}) \, \ln \,(1 - n_{i})\, ]
\end{eqnarray} 
where $m_{C},n_{C}$ is the set of magnetizations of the elementary cube with 
$m_i, n_i$  the average site magnetizations for the $\sigma_i$ and $\tau_i$ spins respectively.
The log terms give the entropy for each of the types of spin. 
Minimizing this free energy gives a set of sixteen coupled mean-field equations
\begin{eqnarray}
\frac{\partial\phi(m_{C}, n_{C})}{ \partial m_{i}}_{(i=1 {\ldots} 8)} &=& 0  \nonumber \\ 
\frac{\partial\phi(m_{C},n_{C})}{ \partial n_{i}}_{(i=1 {\ldots} 8)} &=& 0
\end{eqnarray}
(one for each corner of the cube and spin type)
rather than the familiar single mean field equation for the standard nearest neighbour Ising
action.
The resulting equations are all of the form
\begin{eqnarray}
m_{1}&=& \tanh[\beta  (m_{4} + m_{2} \, n_{1} \, n_{2}) ]\nonumber \\
     &  \vdots& \nonumber \\
m_{8}&=& \tanh[\beta (m_{5} + m_{7} \, n_{7} \, n_{8} ) ]  \\
     &  \vdots& \nonumber \\
n_{1}&=& \tanh[\beta  (n_{5} + n_{2} \, m_{1} \, m_{2}) ]\nonumber \\
     &  \vdots& \nonumber \\
n_{8}&=& \tanh[\beta (n_{4} + n_{7} \,m_{7} \, m_{8}) ] \nonumber
\label{e2a}
\end{eqnarray}  
where we have labelled the magnetizations on a face of the cube counter-clockwise $1 \ldots 4$ 
and similarly for the opposing face $5 \ldots 8$,
as shown in Fig.~(\ref{Fig1}).
If we solve these equations iteratively at different temperatures we arrive at 
zeroes for a paramagnetic phase or various combinations
of $\pm 1$ for the magnetized phases on the 
eight cube vertices. 

A potential problem with an iterative scheme to solve the system of mean field equations
\begin{eqnarray}
m_i^{(k+1)} = f_i[m^k, n^k] \, , \; \;  
n_i^{(k+1)} = f_i[m^k, n^k] \, ,
\end{eqnarray}
is that it might fail to converge if an eigenvalue of
$ \partial m^{(k+1)}_i / \partial m^k_j$ or $ \partial n^{(k+1)}_i / \partial n^k_j$
is less than $-1$ \cite{2}. This is easily remedied by modifying the equations
to 
\begin{eqnarray}
m_i^{(k+1)} &=&  { \left( f_i[m^k, n^k] + \alpha m^k_i \right) \over 1 + \alpha} \nonumber \\
n_i^{(k+1)} &=&  { \left( f_i[m^k, n^k] + \alpha n^k_i \right) \over 1 + \alpha}
\end{eqnarray}
for a suitable $\alpha$, and we have employed this here to ensure stability.

The mean field solution
is then given by gluing together the elementary cubes consistently
to tile the complete lattice, in the manner
of the ground state discussion. 
In the limit $\beta \to \infty$ the mean field equations of equ.(\ref{e2a}) become
the system of equations
\begin{eqnarray}
m_{1}&=& \textrm{sgn} [m_{4} + m_{2} \, n_{1} \, n_{2}]\nonumber \\
     &  \vdots& \nonumber \\
m_{8}&=& \textrm{sgn}[m_{5} + m_{7} \, n_{7} \, n_{8} ]  \\
     &  \vdots& \nonumber \\
n_{1}&=& \textrm{sgn}[n_{5} + n_{2} \, m_{1} \, m_{2}]\nonumber \\
     &  \vdots& \nonumber \\
n_{8}&=& \textrm{sgn}[n_{4} + n_{7} \,m_{7} \, m_{8}] \nonumber
\label{e2b}
\end{eqnarray}
which are, as they should be, compatible with the various ground state structures shown in Fig.~(\ref{ground}). 
Solving the mean field equations numerically finds a (single) transition at $\beta \sim 0.83$ from a paramagnetic state to one of the possible ground states. If the iteration is seeded with spin values close to $\pm 1$ one of the ferromagnetic states is picked.

It would be interesting to refine the mean field solution further by employing a cluster variational
approximation, which essentially amounts to ``improving'' the entropy term and can be combined with Pad\'e
approximant methods to obtain quite accurate critical exponent estimates \cite{11}.
This has been done  successfully for the original Gonihedric model \cite{12}, but we do not pursue this further here, turning instead to Monte Carlo simulations
to sketch out the phase diagram of the model.

\section{Some (Modest) Monte Carlo}

\begin{figure}[h]
\centering
\includegraphics[height=7cm]{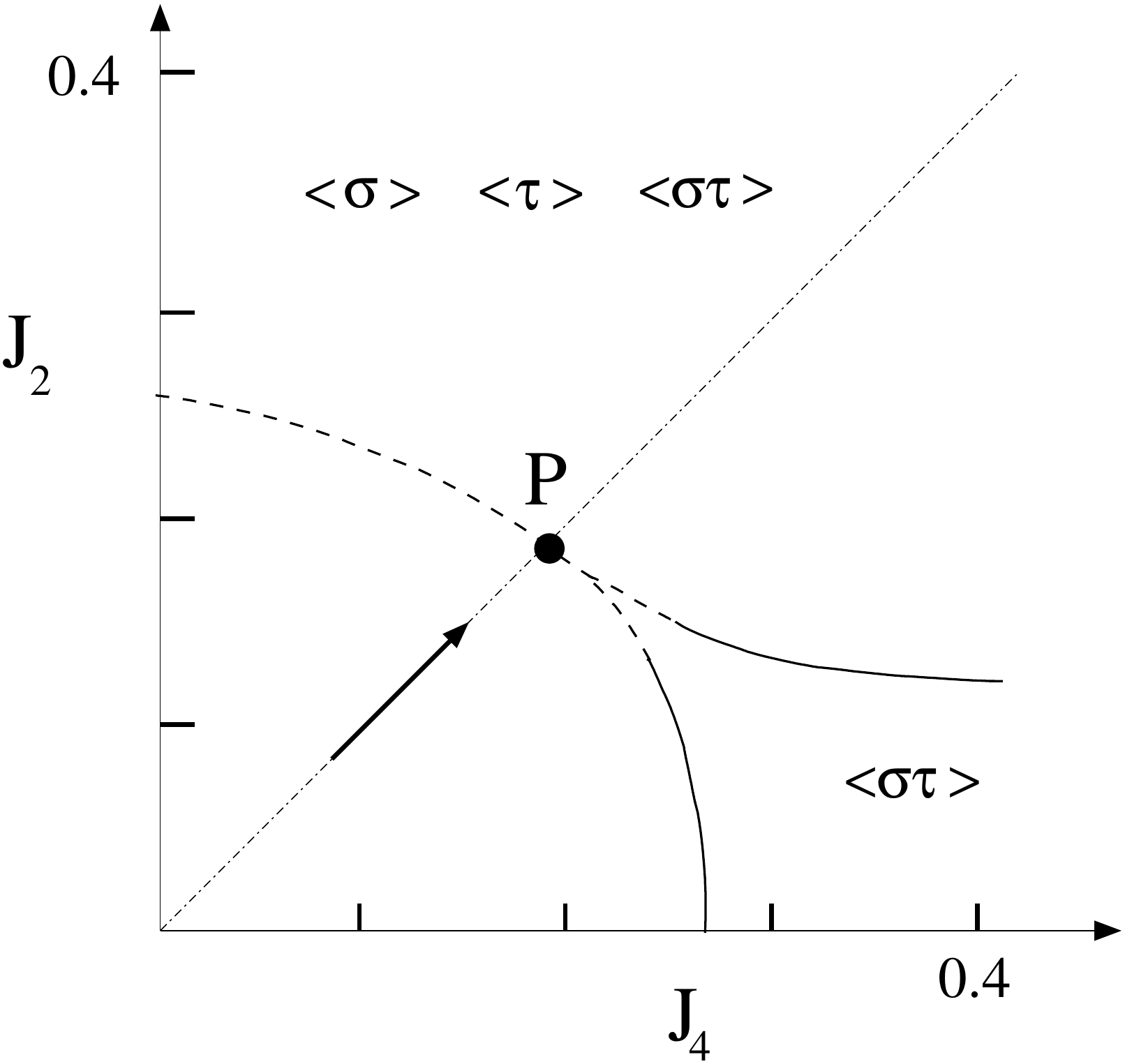}
\caption{A schematic drawing of the phase diagram of the {\it isotropic} 3D Ashkin-Teller model
for positive two-spin, $J_2$, and four-spin, $J_4$, couplings. The 4-state Potts point on
the $J_2=J_4$ line is marked as {\bf P}. The indicated order parameters are non-zero in the phases shown on the diagram, and first order transition lines are shown as dashed lines, second order lines as solid. The paramagnetic phase surrounds the origin and the effect of decreasing the temperature along the $J_2=J_4$ line is indicated by an arrow.}
\label{AT3D}
\end{figure}

For comparison purposes the phase diagram for positive couplings for an isotropically coupled Ashkin-Teller model  is shown
schematically in Fig.(\ref{AT3D}) where $J_2$ is the two-spin and $J_4$ is the four-spin coupling \cite{7,8}. Decreasing the temperature along the $J_2=J_4$ line moves from a paramagnetic phase to a phase in which all of $\langle \sigma \rangle, \langle \tau \rangle$
and $\langle \sigma \tau \rangle$ (``polarization'') are non-zero, which is sometimes called the Baxter phase. The transition takes place at the four state
Potts point and is thus first order, as we saw from a direct rewriting of the Hamiltonian in equ.(\ref{4Potts}).
We are principally interested in determining the nature of the transition in the dual Gonihedric model here, for comparison with the phase diagram of the isotropic Ashkin-Teller model in Fig.~(\ref{AT3D}) rather than carrying out a high accuracy scaling analysis, so we use   relatively modest lattice sizes and statistics in our simulations and employ a simple Metropolis update. Lattices of size
$L=10,12,14,16,18$ and $20$ with periodic boundary conditions for both the $\sigma$ and $\tau$ spins were simulated
using both hot and cold starts at various temperatures. Following a suitable number of thermalization sweeps determined by the energy autocorrelation time, $10^7$ measurement sweeps were carried out at each lattice size for each temperature simulated.

An estimate for the phase transition point of the original plaquette $\kappa=0$ Gonihedric model is the value in \cite{13} which takes account of the (effectively)
fixed boundary conditions employed in the simulations there to fit to a suitable scaling form 
with the correct leading and subleading finite size corrections in such a case,
$\beta_c (L) = \beta_c + a_1/ L + a_2 / L^2$.
This found $\beta_c = 0.54757(63)$. Allowing for factors of 2 in the coupling definitions, an estimate for the dual transition temperature $\beta_c^*$ is then given by
the standard formula $\beta^*_c = -\ln [ \tanh ( \beta_c/2) ] = 1.32$. 

\begin{figure}[h]
\centering
\includegraphics[height=7cm]{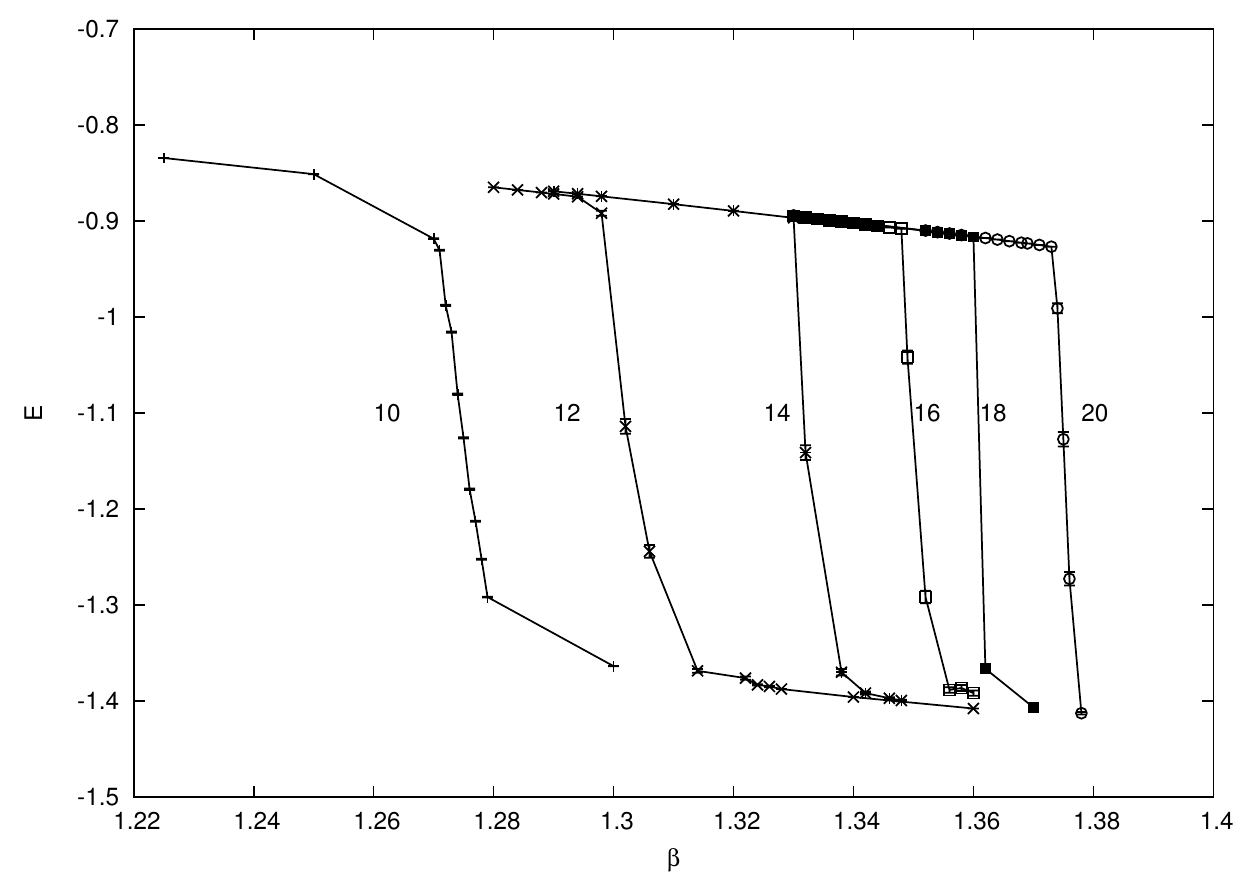}
\caption{The energy for variously sized lattices ranging from $10^3$ to $20^3$ from left to right. The lines joining the data points are drawn to guide the eye and {\it hot} starts have been used in all the simulations.}
\label{E0} 
\end{figure}

A plot of the energy is shown for various lattice sizes in Fig.~(\ref{E0}) where there is clearly a sharp drop in the region of $\beta \sim 1.38$, somewhat higher than the estimate from the dual transition temperature.
We can get a rough idea of the
energy autocorrelation time $\tau_e$ in the simulations by comparing the naive estimate for the variance (i.e. specific heat)
\begin{equation}
	\epsilon^2_{naive} = \sum_{j=1}^{n_m} \frac{\left( E_j -  \langle E \rangle  \right)^2}{n_m -1}
\end{equation}
where $n_m$ is the number of measurements carried out,
with a jack-knifed estimate using binned data $\epsilon_{JK}$. The two are related by
\begin{equation}
	\epsilon_{JK} = \sqrt{\frac{2 \tau_e}{n_m}} \epsilon_{naive}.
\end{equation}
Away from the transition point we find $\tau_e \sim 1$ but large values of $\tau_e \sim 10^3$ appear in its vicinity.

\begin{figure}[h]
\centering
\includegraphics[height=7cm]{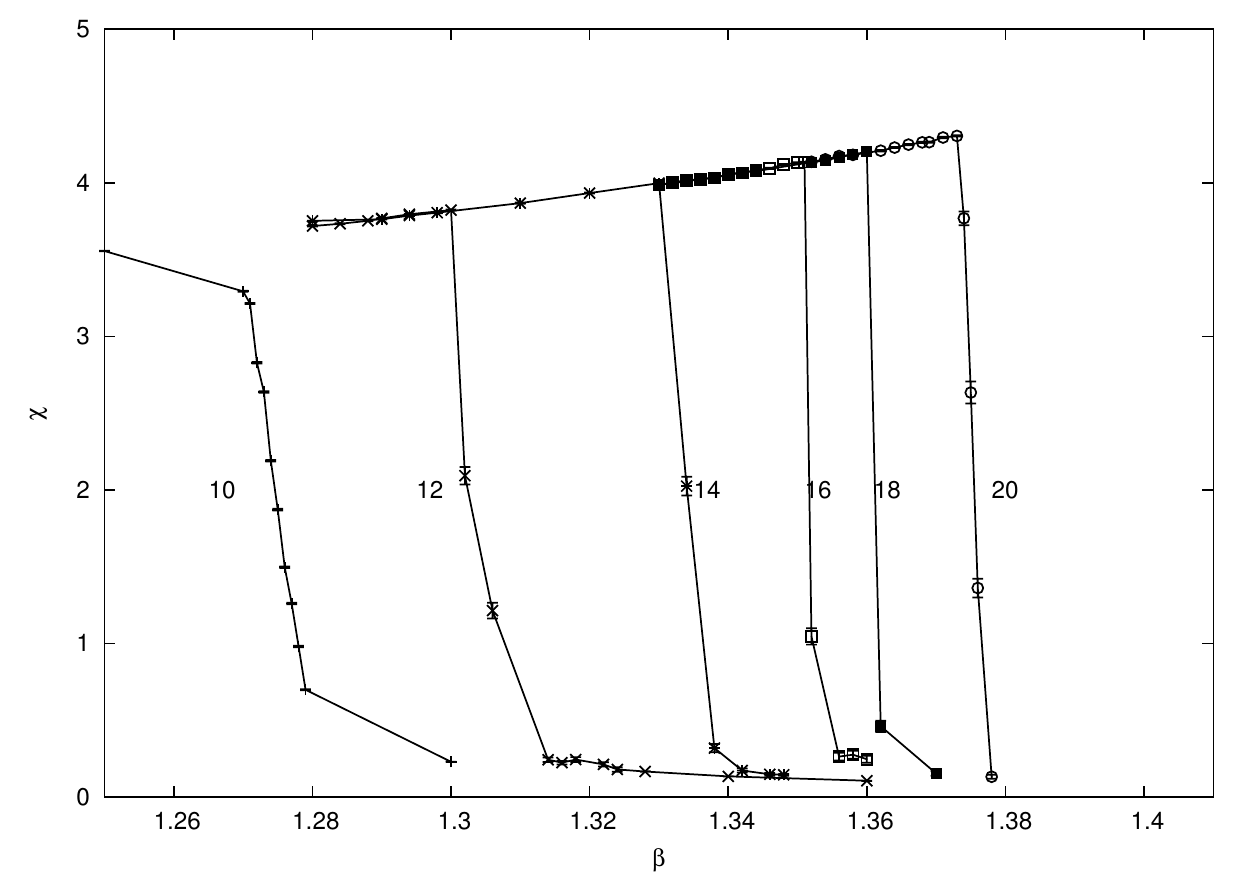}
\caption{The polarization susceptibility, $\chi$ for a $10^3$ to $20^3$ lattices, showing a sharp drop near the pseudo-critical point
in the region of $\beta=1.375$ for the various lattice sizes. As for the energy in Fig.~(\ref{E0}) the lines between the data
points are drawn to guide the eye and hot starts have been used.}
\label{Susp} 
\end{figure}

There is no signal of the observed phase transition in any of the standard magnetic order parameters $\langle \sigma \rangle, \langle \tau \rangle$
and $\langle \sigma \tau \rangle$.
This suggests that the degeneracy observed in the ground state features at finite temperatures as well.
The susceptibility for both the individual $\sigma$ and $\tau$ spins and the polarization $\sigma \tau$ is, however,
non-zero and {\it does} show a signal at the phase transition point where, like the energy, it drops sharply. In Fig.~(\ref{Susp}) we plot
the polarization susceptibility for  $10^3 - 20^3$ lattices. 
The behaviour of the individual spin susceptibilities
is very similar, with that for both the $\sigma$ and $\tau$ spins showing a sharp drop from the high temperature phase to a much lower value at the same point. 

\begin{figure}[h]
\centering
\includegraphics[height=5cm]{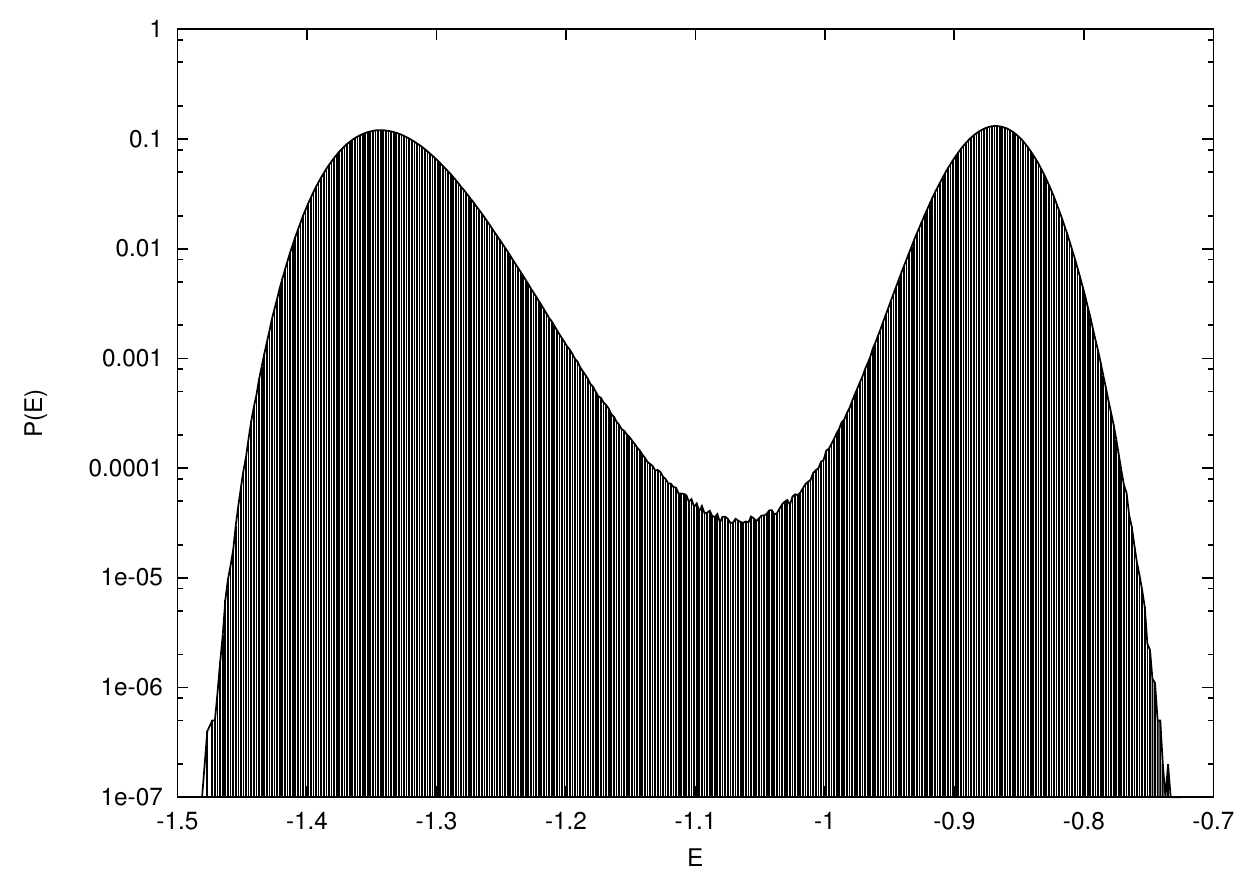}
\caption{The energy histogram from a simulation with $10^7$ sweeps on a $10^3$ lattice near the finite size transition point (in this case
$\beta=1.275$). $P(E)$ is shown on a logarithmic scale.}
\label{bimodal} 
\end{figure}

A good indicator of a first order transition  is a bi-modal energy distribution at the  transition point, so we also histogrammed the energy during the simulations.
Looking at an energy histogram from a simulation sufficiently near the finite size pseudo-critical temperature should display a two-peak structure for  a first order transition.  A typical example for a $10^3$ lattice close to its finite size pseudo-critical temperature at $\beta_c = 1.275$ is shown in Fig.~(\ref{bimodal}), where we have histogrammed the $10^7$ measurements 
of the energy which were carried out after each full lattice sweep of the $\sigma$ and $\tau$ spins.
The double peak structure expected of a first order transition is clearly visible.
A direct consequence of the bi-modal energy distribution near criticality is a non-trivial limit for Binder's energy cumulant. This is defined as
\begin{equation}
	U_E = 1 - \frac{\langle E^4 \rangle}{3  \langle E^2 \rangle^2}
\end{equation}
which approaches $2/3$ at a second order transition point and a non-trivial limit at a first order point. We observe a non-trivial minimum value of $U_E$ of $0.60(2)$, which varies little across the lattice sizes simulated.

Similarly, the $\beta$ value of the minimum of $U_E$ on an $L^3$ lattice, $\beta_{min}(L)$, is expected to scale as $\beta_{min} (L) = \beta_c - O(1 / L^3)$ for  a first order transition.
If we plot the estimated minima positions for the various lattice sizes against $1/L^3$ we get a reasonable fit to this behaviour with a value of $\beta_c \sim 1.388(4)$ and a 
$\chi^2_{dof}$ of $1.34$ when the smallest lattice size of $10^3$ is dropped from the fits.
\begin{figure}[h]
\centering
\includegraphics[height=7cm]{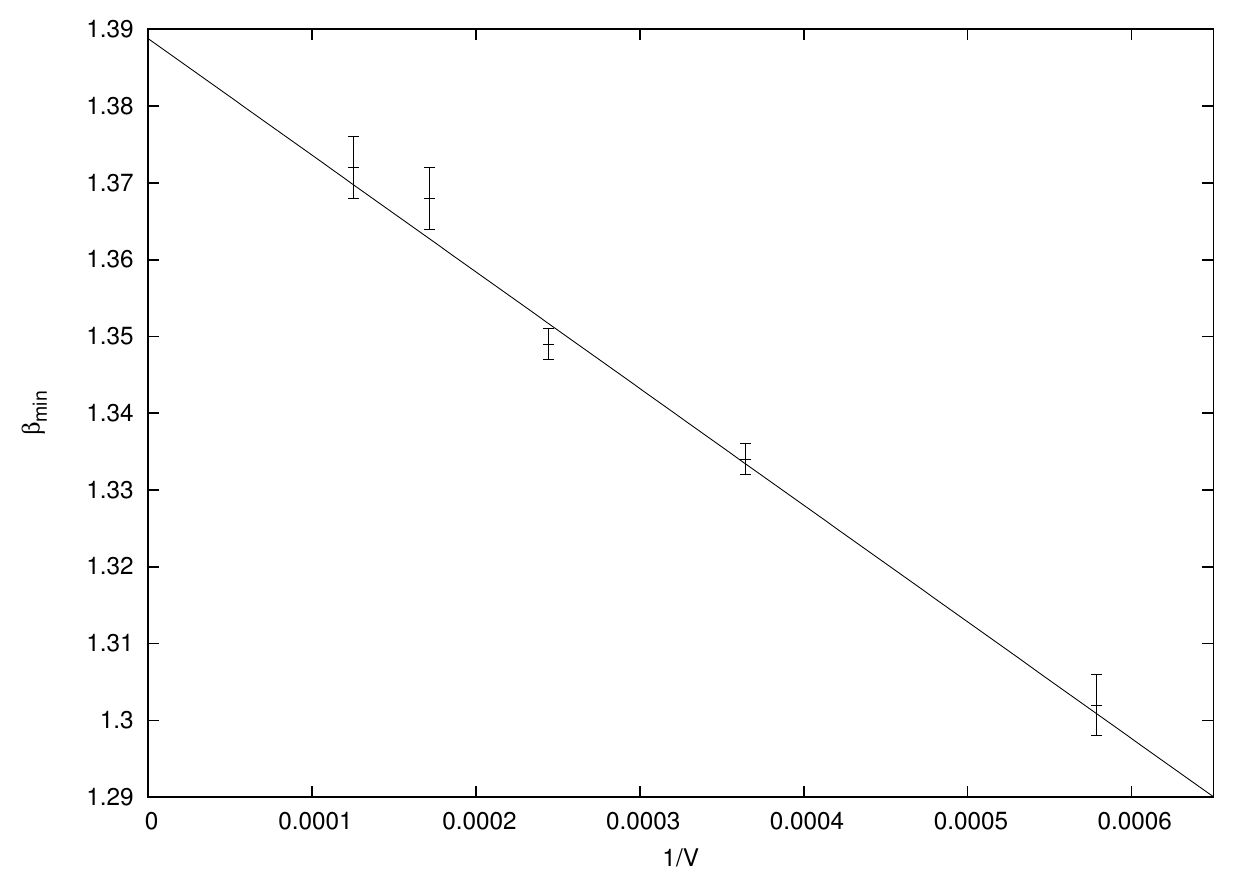}
\caption{The scaling of the position of the minimum of the Binder energy cumulant against the inverse volume.}
\label{E4E2_min} 
\end{figure}
The estimated value of $\beta_c$ from this procedure is consistent with the behaviour of the energy and susceptibility jumps
but, as we have already noted, it is somewhat higher than the dualized value calculated from from the measurements in \cite{13}.  

\begin{figure}[h]
\centering
\includegraphics[height=7cm]{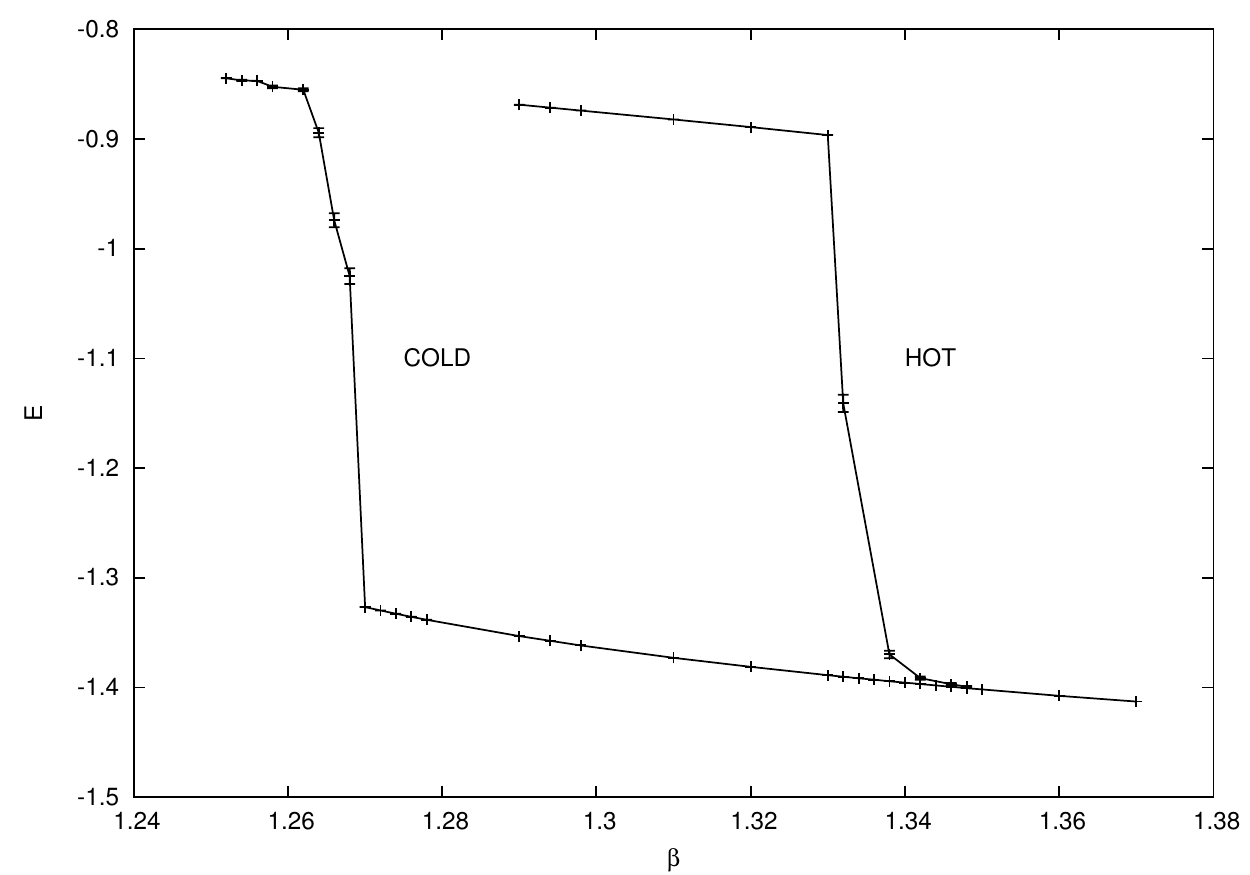}
\caption{Measurements of the energy for hot and cold starts in the region of the transition point on a $14^3$ lattice.}
\label{hotcold14} 
\end{figure}

There is strong hysteresis around the transition point which may affect the accuracy of any such estimates. We show the
result of using both ordered (cold)
and disordered (hot) starting configurations for a $14^3$ lattice in Fig.~(\ref{hotcold14}) with, in both cases, relaxation times of $10^4$ sweeps followed by $10^7$ measurement sweeps.
This behaviour is again strikingly similar to that seen in the original plaquette Hamiltonian \cite{5a}.

\section{Some (Very Modest) Dynamics}
   
An intriguing feature of the original $\kappa=0$ Gonihedric model is its highly non-trivial non-equilibrium behaviour, including a region of metastability around its first order transition point and a dynamical transition   
which displays many glassy characteristics. We have seen that the dual Gonihedric model's equilibrium phase diagram
is similar to the original, so it is natural to inquire whether this similarity also holds for  dynamical behaviour. 

\begin{figure}[h]
\centering
\includegraphics[height=7cm]{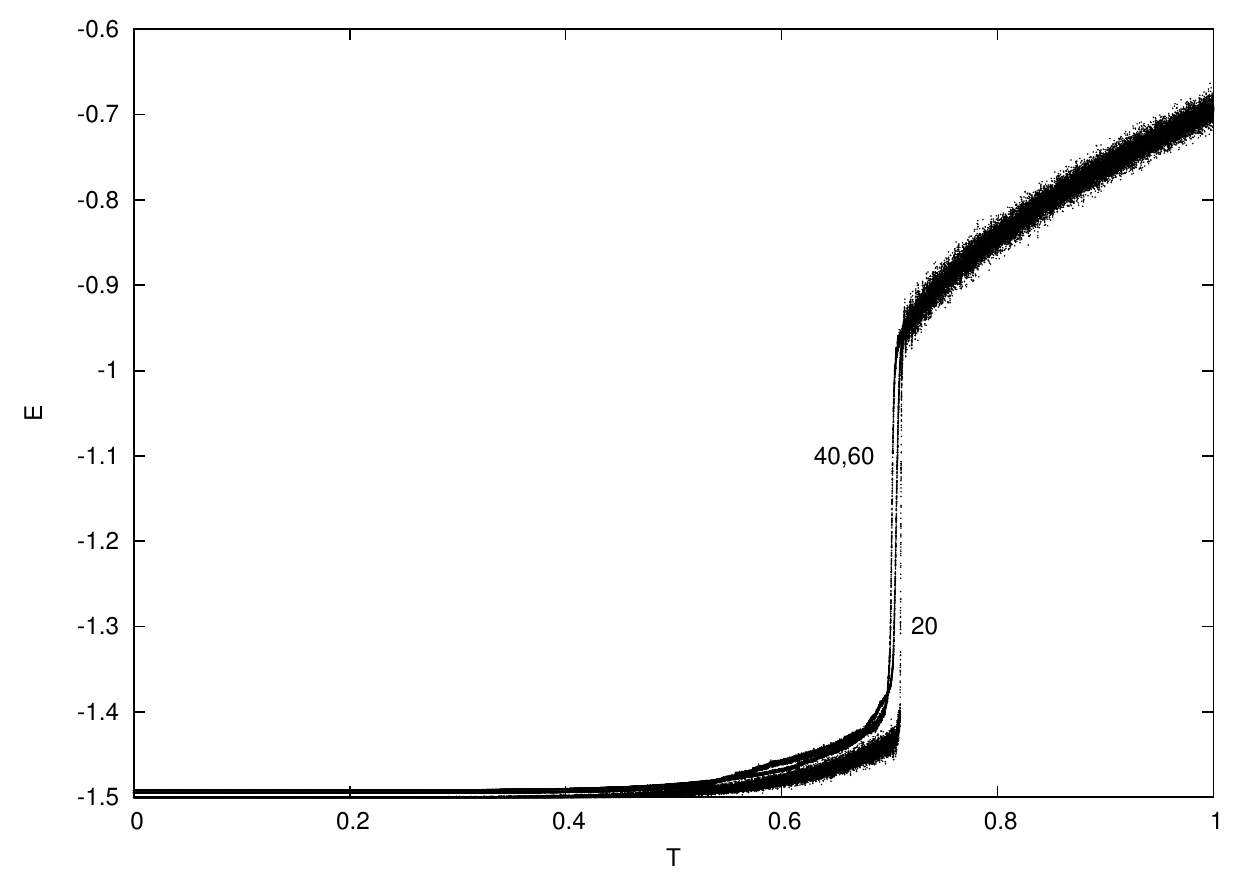}
\caption{The time series of energy measurements obtained from cooling $20^3, 40^3$ and $60^3$ lattices from a hot start
at a rate of $\delta T = 0.00001$ per sweep.}
\label{r00001} 
\end{figure}

\begin{figure}[h]
\centering
\includegraphics[height=7cm]{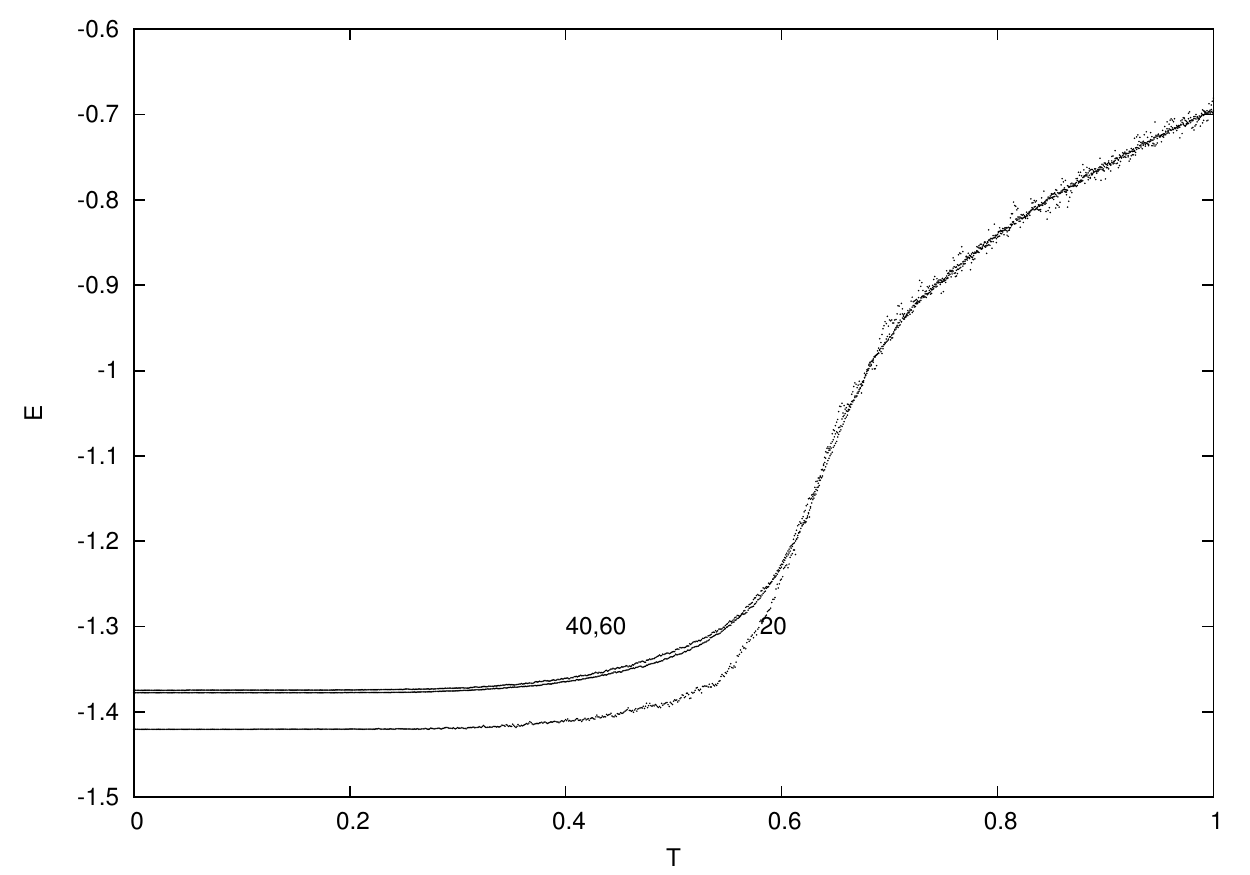}
\caption{The time series of energy measurements obtained from cooling $20^3, 40^3$ and $60^3$ lattices from a hot start
at a rate of $\delta T = 0.001$ per sweep.}
\label{r001} 
\end{figure}

As a start in this direction, we cooled differently sized lattices at various rates from disordered (hot, ``liquid'') starts
in order to see if there was any evidence of the potentially glassy behaviour seen with the plaquette Hamiltonian.
We cooled lattices of size $20^3, 40^3$ and $60^3$ starting with a disordered configuration at a temperature of $T=3$. From Figs.~(\ref{r00001},\ref{r001}) it is clear that there is little difference between the $40^3$ and $60^3$ lattice results, while the $20^3$ lattices may still be subject to stronger finite size effects. With regards to the numerical estimates of critical temperatures, this also suggests that the equilibrium simulations in the previous section may have been carried out on rather small lattices.

In Fig.~(\ref{r00001}) we can see that with a slow cooling rate of $\delta T = 0.00001$ per sweep, the systems still relax 
to a (ground) state with $E = -1.5$. The jump in the energy at the phase transition seen in the time series at $T \sim 0.72$ on the larger lattice sizes  is consistent  
with the estimate of $\beta_c = 1.388(4)$ obtained from extrapolating the Binder cumulant values.
Fig.~(\ref{r001}), where a faster cooling rate of $\delta T = 0.001$ per sweep is employed, is perhaps more interesting. Once again there is little difference between the $40^3$ and $60^3$ lattices,
but this time they do {\it not} relax to the ground state energy of $E=-1.5$, but are trapped at a higher value.
Similar results in the plaquette model \cite{5a,5} were taken as an indicator of possible glassy behaviour, though
there is evidence that what is seen is an echo of a mean-field spinodal point \cite{5b} at which the supercooled high-temperature (``liquid'') phase  becomes physically irrelevant.

\section{Discussion}
We have studied the dual of the $\kappa=0$ Gonihedric Ising model in three dimensions, which may 
be formulated as an anisotropically
coupled Ashkin-Teller model. We  noted that this dual Gonihedric model displays clear signals of a first order transition, such as a bimodal energy histogram and a non-trivial limit for Binder's energy cumulant (like the isotropically coupled equivalent),  but it
has a highly degenerate ground state ({\it un}-like the isotropically coupled equivalent). 

The magnetic order parameters $\langle \sigma \rangle ,  \langle \tau \rangle , \langle \sigma \tau \rangle$ show no strong signal at the phase transition point which is, however, clearly visible in the energy and various susceptibilities 
as well as the  Binder (energy) cumulant. The absence of conventional ferromagnetic order 
suggests that the degenerate ground state structure persists to finite temperatures, as is the case for the $\kappa=0$ Gonihedric model. It would be an interesting exercise to carry out a low temperature expansion of the dual Hamiltonian to verify this  by comparing the energy of the  states with flipped spin planes to a ferromagnetic reference state in the manner of \cite{10}.

Another aspect which merits investigation is the crossover to the isotropic model, 
which has a much simpler ground state structure and known, simple order parameters for the low temperature phase(s). In any such endeavours the form of the anisotropic action suggests that it might be more amenable to a cluster simulation than the original plaquette action. The first order nature of the transition means that this would not confer such great advantages over local updates as in the case of continuous transitions, though  cluster updates could be employed in conjunction with multihistogramming methods of various sorts for maximum numerical efficiency.  

To investigate the non-equilibrium behaviour of the model, however,  Metropolis (or other local) dynamics should be employed.
We have made a start in this by conducting some cooling experiments which show that the phenomenology of the dual Hamiltonian appears to be remarkably similar to that of the original $\kappa=0$ plaquette Hamiltonian. Under very slow cooling the ground state energy is achieved, but faster cooling appears to trap the system in a higher energy state. More extensive simulations along the lines of those conducted in \cite{5e,5b} for the plaquette model and
the the coupled two layer system (CTLS) would be useful to discern whether the  ``bubbling  and 
coarsening'' scenario posited there for the low temperature behaviour also applied in the case of the dual Gonihedric model, or whether more conventional coarsening dynamics was seen. 

If the low temperature behaviour does, indeed, display (pseudo-)glassy characteristics it would also be useful to elucidate the nature of the self-induced frustration which is presumably causing it. For both the plaquette Gonihedric model and the CTLS multi-spin interactions appear to play a vital role. In view of the simplicity of the Hamiltonian, the dual Gonihedric model might also provide a further test case in which to explore the approach of \cite{14}, which links dynamical, glassy  behaviour in classical systems to quantum phase transitions.

\section{Acknowledgements}
The work of R. P. K. C. M. Ranasinghe was supported by a Commonwealth Academic Fellowship {\bf LKCF-2010-11}. D. A. Johnston would like to thank W. Janke for useful discussions and comments.


\end{document}